\documentclass[prl,aps,twocolumn,amsmath,amssymb]{revtex4}

\usepackage{graphicx}
\usepackage{dcolumn}
\usepackage{bm}

\begin{document}

\preprint{APS/123-QED}

\title{Multiple layer local oxidation\\
for fabricating semiconductor nanostructures}

\author{M. Sigrist$^1$, A. Fuhrer$^1$, T. Ihn$^1$, K. Ensslin$^1$, D. C. Driscoll,$^2$ and A. C. Gossard$^2$}
\affiliation{
$^1$Solid State Physics, ETH Z\"urich, 8093 Z\"urich, Switzerland\\
$^2$ Materials Department, University of California, Santa Barbara, CA 93106, USA\\}

\date{\today}

\begin{abstract}
Coupled semiconductor nanostructures with a high degree of
tunability are fabricated using local oxidation with a scanning
force microscope. Direct oxidation of the GaAs surface of a
Ga[Al]As heterostructure containing a shallow two-dimensional
electron gas is combined with the local oxidation of a thin
titanium film evaporated on top. A four-terminal quantum dot and a
double quantum dot system with integrated charge readout are
realized. The structures are tunable via in-plane gates formed by
isolated regions in the electron gas and by mutually isolated
regions of the Ti film acting as top gates. Coulomb blockade
experiments demonstrate the high quality of this fabrication
process.
\end{abstract}

\maketitle Optical lithography and electron beam lithography are
the standard techniques to pattern tunable semiconductor
nanostructures. A number of patterning techniques based on
scanning force microscopes (SFMs) have been developed
~\cite{Dagata90,Wendel94,Minne95,Avouris97,Shirakashi98,Rosa98,Rokhinson02,Ishii95,Held97,Luscher99,Keyser00,Fuhrer02,Neumutudi02,Sugimura93,Irmer97,Held99,Matsumoto96,Held00,Rogge03}.
An especially useful technique is to oxidize substrates locally by
applying a negative voltage between the SFM tip and the substrate.
Sophisticated nanostructures can be fabricated, for example, on
shallow Ga[Al]As
heterostructures~\cite{Ishii95,Held97,Luscher99,Keyser00,Fuhrer02,Neumutudi02}.
The two-dimensional electron gas is depleted below oxide lines
leading to mutually isolated regions of electron gas. The main
experimental parameters influencing the oxidation process are the
relative humidity of the environment, leading to a thin water film
on the substrate surface, and the magnitude of voltage applied to
the tip.

\begin{figure}[bthp]
\includegraphics[width=7cm]{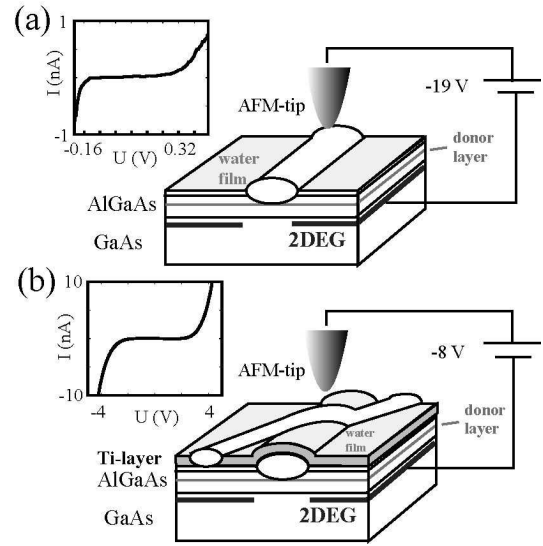}
\caption{\label{schematic} (a) Schematic of the GaAs-oxidation
step. Local oxidation of the GaAs surface depletes the electron
gas locally below the line. The inset shows the current between
separated regions of the 2DEG measured at 4.2~K. (b) A thin Ti
film is evaporated and oxidized using a voltage biased SFM-tip.
The inset shows the current between separated regions of the film
measured at a temperature of 4.2~K.}
\end{figure}

A similar oxidation technique has been used to divide thin
Titanium films into areas mutually isolated by oxide
lines~\cite{Sugimura93,Irmer97,Held99}. Metallic single-electron
transistors fabricated with this technique have been reported to
show the Coulomb-blockade effect~\cite{Matsumoto96}. Local
titanium oxidation has also been used on Ga[Al]As heterostructures
with a thin Ti top gate film for defining self-aligned
split-gates. On the application of appropriate gate voltages, a
quantum point contact (QPC) was formed showing clear conductance
quantization~\cite{Held97,Held00}.

As the size of nanostructures to be defined with these techniques
becomes comparable with the full width of the oxide lines of
typically 150~nm and the desired degree of control of individual
parts of the structure requires more and more gate electrodes,
neither local oxidation of the GaAs surface nor that of a titanium
top gate alone meets the needs. The limits are of geometrical
nature and imposed by the restriction of patterning only within
the two-dimensional plane of the Ti or GaAs surface. Therefore, a
patterning technique is desirable that makes the third spatial
dimension accessible. One option is the use of local oxidation on
the surface of the semiconductor, followed by the deposition of
top gates defined with electron beam lithography
techniques~\cite{Rogge03}. In this approach the alignment of the
top gates with SFM-defined structures has to be achieved using
another electron beam lithography step prior to the local
oxidation for the definition of markers on the surface.

In this letter, we demonstrate an approach where the alignment of
the top gate structures with the previous oxide structure in GaAs
is achieved using a SFM. Our method combines the direct oxidation
of a Ga[Al]As heterostructure (first step) with the subsequent
evaporation of a thin titanium gate (second step) which is then
patterned using the local oxidation technique with the SFM (third
step). We demonstrate the fabrication technique with the
realization of a four-terminal quantum dot (sample A) and a
double-quantum dot with integrated charge read-out (sample B). The
occurrence of Coulomb blockade demonstrates the high quality and
stability of the structures. The tuning possibilities of the
nanostructures are significantly increased by the metallic Ti
gates on top and their alignment with the nanostructure is
straightforward.

The fabrication process is based on high-quality Ga[Al]As
heterostructures containing a two-dimensional electron gas (2DEG)
34 nm and a back gate 1.4~$\mu$m below the surface. A $22\times
22$~$\mu$m$^{2}$ mesa with Ohmic contacts and metallic top gate
fingers for contacting the Ti film are defined by
photolithography.

In the first step the nanostructure is defined by direct oxidation
of the GaAs surface using the voltage-biased tip of a
SFM in a humidity-controlled environment. The
resulting oxide lines deplete the electron gas below
[Fig.~\ref{schematic}~(a)]. Typical break down voltages across an
oxide line in the 2DEG are a few hundred mV at liquid He
temperatures [an example is shown in the inset of
Fig.~\ref{schematic}~(a)]. Details of this fabrication step are
described in Refs.~\onlinecite{Fuhrer02} and~\onlinecite{Held99}.
In the second step (using electron beam lithography), we define a
$22\times 22$~$\mu$m$^{2}$ mask overlapping all the top gate
fingers and evaporate a 6-7~nm thick titanium film. The metal
layer on the resist is stripped with a lift-off process. The
resistivity of the remaining Ti film is of the order of
10~k$\Omega$ and the surface roughness is 1-2~nm peak-to-peak
which is important for further structuring. A Schottky barrier
forms between the heterostructure and the Ti film allowing its use
as a top gate.

In the third step, the top gate is split into regions of dedicated
functionality by oxidizing the Ti film locally, again with the
voltage-biased tip of the SFM as shown in Fig.
\ref{schematic}~(b). Since we have not found a clear selectivity
between the oxidation process of GaAs and that of Ti for any
humidity or tip bias parameters, the writing parameters are chosen
carefully, such that the 2DEG density is not measurably reduced by
the Ti lithography on top of the GaAs-structure. For the Ti
oxidation process, the SFM is operating in tapping mode controlled
by a standard feedback loop. The relative humidity at room
temperature is kept constant at ($43\pm0.5$) percent. A voltage of
-7 to -9~V between SFM tip and Ti film oxidizes the Ti film
locally without affecting the GaAs surface significantly. A
control parameter for this process is the (averaged) height of the
Ti oxide lines which further increases when the GaAs below the Ti
film starts to be oxidized for voltages below -10~V. We found that
lines of 2-3~nm height and 100~nm width isolate regions in the Ti
film at low temperatures with typical break down voltages of a few
Volts [inset of Fig. \ref{schematic} (b)]. More details about the
local Ti oxidation technique are found in
Refs.~\onlinecite{Irmer97,Held97}.

\begin{figure}[bthp]
\includegraphics[width=7cm]{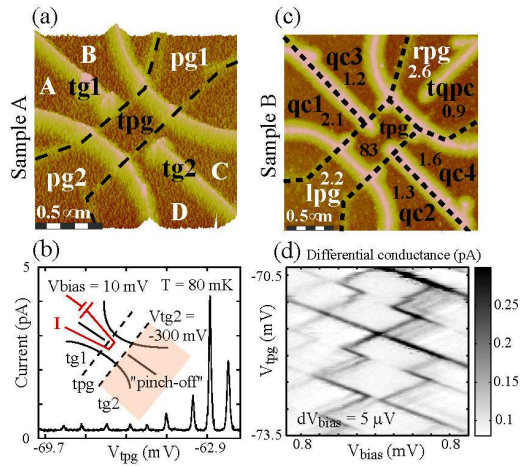}
\caption{\label{ms2p1} (a) SFM-micrograph of sample A. The 2DEG is
depleted below the oxide lines on the GaAs surface (grey). The Ti
film is structured into three isolating parts by Ti-oxide lines
(dashed lines). (b) Conductance is measured across the quantum dot
of sample A as a function of the top plunger gate (tpg) at 80~mK.
(c) SFM-micrograph of sample B. Dashed lines indicate the second
pattern in the Ti film with its six top gate segments. The lever
arm of the gates on the quantum dot are estimated and quoted in
percent (white numbers). (d) Differential conductance of the lower
left dot of sample B is measured as a function of voltage bias and
top plunger gate voltage.}
\end{figure}

Sample A is shown in Fig.~\ref{ms2p1}~(a). A quantum dot (QD)
structure connected to four distinct leads via quantum point
contacts (QPCs) is defined by local oxidation of the GaAs surface
(grey lines). This structure with only two in-plane gates does not
allow individual tuning of the four QPCs and the dot. A thin Ti
film is evaporated onto the structure and segmented by two oxide
lines (dashed lines). The QPCs are labelled A,B,C,D (white). The
Ti top gate regions are named (black letters) top gate~1~(tg1),
top plunger gate~(tpg) and top gate~2~(tg2). The in-plane gates
formed by regions of the 2DEG next to the QD are labelled (white
letters) plunger gate~1 and~2 (pg1 and pg2). Within measurement
accuracy we do not observe a change in the QPC conductance at
4.2~K after writing oxide lines in the Ti film on top. The three
electrically separated regions of the top gate together with the
two in-plane gates give five independent control parameters that
can be used to tune the four QPCs and the dot. In addition, the
global electron density of the structure can be tuned with the
back gate. The separation of the two Ti-oxide lines is less than
290~nm peak-to-peak and and the Ohmic resistance of the top gate
tpg bridging the structure is only a few k$\Omega$s larger than
that of the unpatterned Ti film. In order to demonstrate the
functionality of the top and in-plane gates we have measured their
action on the conductance of selected QPCs relative to the action
of the back gate at a temperature of 4.2~K.
Conductance is measured from
QPC A to QPC B by pinching off the two other QPCs with top gate~2. Clear Coulomb blockade is observed  (see Fig.~\ref{ms2p1} (b)) as a
function of the top plunger gate at an electronic temperature of
about 80~mK.

Sample B is a double QD structure as illustrated in
Fig.~\ref{ms2p1}~(c). The dashed black lines indicate the oxide
lines in the Ti film, the thick bright lines are those on the GaAs
surface. Each of the four QPCs connecting the dots to the leads
can be controlled by a separate top gate (qc1\ldots qc4). In
addition, two in-plane gates (lpg and rpg) act as plunger gates
for the quantum dots. The coupling between the dots can be tuned
by the top plunger gate (tpg). The detector QPC flanking the top
right QD forms an integrated charge read-out~\cite{Field93}. It is
tunable with another top gate region (tqpc).

\begin{figure}[tbhp]
\includegraphics[width=7cm]{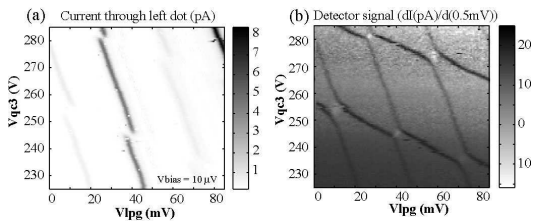}
\caption{\label{ms2p4} (a) Conductance through the lower left dot
of Fig.~\ref{ms2p1}~(c) is measured as a function of the left
in-plane gate $V_{lpg}$ and the top gate segments
$V_{qc3}=V_{qc4}+130~mV$. (b) The full charge stability diagram of
the double dot system is measured in the QPC detector signal.}
\end{figure}

Coulomb diamonds are shown in Fig.~\ref{ms2p1}~(d) as differential
conductance maps measured through the contacts below top gate
areas qc1 and qc2. The charging energy of each quantum dot is
about 0.5~meV. The lever arms of the gates on the lower left dot
in this regime have been measured from Coulomb blockade diamonds
as a function of the top plunger gate (see Fig.~\ref{ms2p1}~(d))
and conductance measurements in the parameter plane of the top
plunger gate and all other gates. Since the sum of all lever arms
is unity, we quote the numbers for individual gates in percent in
Fig.~\ref{ms2p1}~(c). The lever arm of the top plunger gate (tpg)
is significantly larger than that of all other gates. The lever
arms of the quantum point contact top gates (qc1 to qc4) are
comparable to those of the in-plane gates (lpg,rpg). The lever arm
of the top gate above the detector QPC (tqpc) is the smallest.
These findings agree with expectations from the sample geometry.

Figure~\ref{ms2p4}~(a) shows a conductance map for transport through the lower left dot of Fig.~\ref{ms2p1}~(c).  The tunnel couplings of the upper right dot are completely pinched off in this regime of voltages applied to gates qc3 and qc4. The data is taken by using qc3 and qc4 as tuning gates with $V_{qc3}=V_{qc4}+130~mV$. Dark lines correspond to conductance resonances as a function of the two tuning parameters, the left in-plane gate $V_{lpg}$ and the top gate
segments qc3 and qc4.

When the charge detector QPC is tuned into the tunneling regime we are able to detect charging of the individual dots
with single electrons \cite{Field93}. The charge stability diagram of the double
dot system with its characteristic hexagon pattern
~\cite{Livermoore96} can be mapped out as a function of two
tuning gates (see Fig.~\ref{ms2p4}~(b)). The charge detector is closer to the upper right dot in Fig.~\ref{ms2p1}~(c) and therefore shows a stronger signal if this dot is charged.  This gives rise to the lines with the smaller slope in Fig.~\ref{ms2p4}~(b) in agreement with the relative location of this dot to the two sets of tuning gates, respectively. The lines with the larger slope and weaker contrast lie exactly on the conductance maxima presented in Fig.~\ref{ms2p4}~(a) reflecting the charging of the dot in the lower left. This demonstrates individual
control over nanostructured in-plane and top gate electrodes.

The two structures discussed above demonstrate the functionality
of complex coupled nanoscale systems patterned in the Ga[Al]As
system with the layer by layer local oxidation technique. The
alignment of the pattern in the Ti film relative to the oxide
lines on GaAs is straightforward and only limited by the
resolution of the SFM.  Stable Coulomb peaks demonstrate the
high quality of the resulting structures. In addition, these results
enable the fabrication of coupled structures and hold promise for innovative designs of highly tunable semiconductor nanostructures.

The authors are grateful to the Swiss National Science Foundation
(Schweizerischer Nationalfonds) for financial support.



\end{document}